\documentstyle[epsf,preprint,aps]{revtex}
\draft
\preprint{HUTP--98/A009}
\begin{document}
\title{\Large\bf Constraints on Very Light Axions from Cavity
Experiments}
\author{Jihn E. Kim} 
\address{Lyman Laboratory of Physics, Harvard University,
Cambridge, MA 02138, and\\
Department of Physics, Seoul National University,
Seoul 151-742, Korea\footnote{Permanent address}}
\maketitle

\begin{abstract} 
In view of the ongoing galactic (or cosmic) axion detection experiments,
we compare the axion-photon-photon coupling $c_{a\gamma\gamma}$'s 
for various invisible (or very light) axion models. 
\end{abstract}

The $\bar\theta$ parameter of the standard model is naturally
understood in axion models \cite{reviews}. This axion 
interpretation has led to the very light (or invisible) axion's role 
in the galaxy formation \cite{cbound}. If the seed of our galaxy is
indeed the density perturbations due to invisible axions,
the cold dark matter might be these cold axions with $\sim$ 0.3
GeV/cm$^3$ energy density in our galaxy, which for $f\mu$eV axion 
mass corresponds to $\sim 3\times 10^{14}/f$ axions per cm$^3$ 
around us.
 
These ubiquitous axions can be detected using a cavity immersed
in strong magnetic field \cite{sikivie}.  Two groups have already
reported on this type of experiments \cite{cav1,cav2}.  In addition,
$^{139}$La M1 transition has been studied to get a clue on these
galactic axions \cite{min}. These previous experiments have 
given the upper bound on the detection rate, but have not reached
to the level of detecting the galactic axions (or cosmic axions). 

Currently, there are two ongoing experiments, the Kyoto University
experiment on Rydberg atoms \cite{kyoto} and the Lawrence Livermore
National Laboratory (LLNL) experiment \cite{llnl}. In particular,
the sensitivity of LLNL experiment is at the level of
distinguishing several invisible axion models. So far, the
theoretical invisible axion models compared with data are not 
given with proper distinction. Thus it is very important at
this stage to clarify the prediction of the axion-photon-photon
coupling constant $c_{a\gamma\gamma}$ in various invisible (very
light) axion models. 

It is known that the free energy $V$ is minimum at $\bar\theta=0$
in a world without weak CP violation \cite{pq}. If 
$\bar\theta\ne 0$, the QCD term
\begin{equation}
{\bar\theta\over 32\pi^2}
F_{\mu\nu}^a\tilde F^{a\mu\nu}
\end{equation}
violates P and CP symmetry, implying $|\bar\theta|<10^{-9}$
from the neutron electric dipole moment bound. 
This can be understood if we let $\bar\theta$ be a dynamical
variable, i.e.
\begin{equation}
\bar\theta={a\over F_a}
\end{equation}
where $a$ is a pseudoscalar field called axion and $F_a$ is the
axion decay constant. Models with $F_a\gg 250$ GeV give the
so-called {\it invisible axions}, but in view of the 
possible detection of the galactic axions, it is better for
them to be called {\it very light axions}. These axions settle
$\bar\theta$ at $\sim 0$ in an evolving universe even if one starts
from any initial value of $\bar\theta$, due to the potential
of $a$. But introduction of weak CP violation shifts the
minimum position of $\bar\theta$ slightly \cite{georgi} to
$\sim 10^{-17}$ which is far below $10^{-9}$. 

The invisible axions come in three broad categories, depending
on how $a$ arises: (i) pseudo-Goldstone boson \cite{ksvz,dfsz}, 
(ii) fundamental field in string theory \cite{witten}, and 
(iii) composite axions \cite{comp}. Among these, we will
concentrate on the first category, the so-called KSVZ and DFSZ
models.

The calculation of the axion-photon-photon coupling is performed
in two stage, above the chiral symmetry breaking scale and
below the chiral symmetry breaking scale, so that the
coupling is written in the form \cite{kaplan},
\begin{equation}
c_{a\gamma\gamma}=\bar c_{a\gamma\gamma}- {2\over 3}\cdot
{4+Z\over 1+Z}
\end{equation}
where $Z=m_u/m_d$. In any model, the 
chiral symmetry breaking correction is given as the second
term of Eq. (3). The first term of Eq. (3) is given in terms
of the Peccei-Quinn charges of fermions,
\begin{equation}
\bar c_{a\gamma\gamma}={E\over C},\ \ E={\rm Tr}Q_{PQ}Q_{em}^2,
\ \ C\delta_{ab}={\rm Tr}\lambda_a\lambda_b Q_{PQ}
\end{equation}
where $Q_{em}$ is the electric charge operator and 
Tr$\lambda_a\lambda_b={1\over 2}\delta_{ab}$ for the triplet
representation of $SU(3)_c$. The invisible axion
resides mostly in the phase(s) of a complex standard model 
singlet field(s) $\sigma$,
\begin{equation}
\sigma={v+\rho\over\sqrt{2}}e^{ia/F_a}
\end{equation}
where $v$ is the VEV of $\sigma$.
How $\sigma$ couples to quark
fields distinguishes different invisible axion models.
The KSVZ axion couples as
\begin{equation}
{\cal L}=f\bar Q_LQ_R\sigma +{\rm h.c.},
\end{equation}
where $Q$ is a heavy quark, while the DFSZ axion couples as
\begin{equation}
{\cal L}=\lambda\sigma\sigma H_1H_2+\sum_{ij}(f_d^{ij}\bar q^i_Ld_R^jH_1
+f_u^{ij}\bar q_L^iu_R^jH_2)+{\rm h.c.}
\end{equation}
where $H_1$ and $H_2$ are the two Higgs doublets of the
standard model.
These models have $U(1)_{PQ}$ symmetry. The corresponding PQ current 
for the KSVZ axion is
\begin{equation}
J_\mu^{KSVZ}=v\partial_\mu a -{1\over 2}\bar Q\gamma_\mu\gamma_5 Q
\end{equation}
while the current for the DFSZ axion is
\begin{equation}
J_\mu^{DFSZ}\simeq v\partial_\mu a+{x^{-1}\over x+x^{-1}}
\sum_i \bar u^i\gamma_\mu\gamma_5 u^i
+{x\over x+x^{-1}}
\sum_i\bar d^i\gamma_\mu\gamma_5 d^i
+({\rm leptonic\ terms})
\end{equation}
where $x=\langle H_2^0\rangle/\langle H_1^0\rangle=\tan\beta$.
In the DFSZ model, we neglected the small contribution from
two Higgs doublets. The $Q_{PQ}$ is calculated from these
currents. We simplified the models by introducing only one
$\sigma$. 

In the original KSVZ model, we 
introduced only one heavy quark for simplicity.

In the DFSZ model, there is leptonic contribution
in general. If VEV of $H_1$ ($H_2$) gives masses to 
charged leptons, the coefficient of the leptonic current is 
the same as that of $Q_{em}=-1/3$ (2/3) quark. On the other 
hand, if a third Higgs doublet is used to give masses to
charged leptons, the leptonic terms vanish.

In Table 1, various values of $c_{a\gamma\gamma}$ are
presented for the KSVZ and DFSZ models. $Z\simeq 0.6$ is used.
$e_R$ means the electromagnetic charge of the heavy quark color
representation $R$ in units of the positron charge.
In the KSVZ model, a model with $m$ heavy quarks of
$e_3=2/3$ and $n$ heavy quarks of $e_3=-1/3$ is represented
as an $(m,n)$ model. The $(m,m)$ model with any value for
$m$ gives the same result. The $(1,2)$ model is also shown.
In the DFSZ model, $(d^c,e)$ 
unification corresponds to the case where $H_1$ gives mass to
electron, $(u^c,e)$ to the case where $H_2$ does, and nonunification
to the case of a third Higgs doublet. An example of the $(d^c,e)$ 
unification is the familiar $SU(5)$ unification \cite{gg}, 
and an example of $(u^c,e)$ unification is the flipped
$SU(5)$ \cite{flip}. The third case is denoted as nonunification.
Nonunification superstring models obtained considerable 
attention because they have no need for a GUT symmetry breaking 
mechanism \cite{non}. But these nonunification superstring
models can contain $(d^c,e)$ and $(u^c,e)$ models, depending
on how the Higgs doublets couple. Note that $c_{a\gamma\gamma}$
is very sensitive to the electromagnetic charge of the
heavy quark in the KSVZ model and to the ratio of VEV's
of the Higgs doublets in the DFSZ model. Therefore, one can
distinguish different models. 

\vskip 0.5cm
\centerline{Table 1. $c_a\gamma\gamma$ for several KSVZ and DFSZ
models.}
\begin{center}
\begin{tabular}{|cc|cc|}
\hline
\ \ \ \ KSVZ& & DFSZ &\\
$e_R$ & \ \ \ $c_{a\gamma\gamma}$ & $x$ (unif)  & 
\ \ \ $c_{a\gamma\gamma}$\\
\hline
$e_R=0$ & \ \ \ \ --1.92 & any ($d^c,e$) & \ \ \ \ 0.75\\
$e_3=-1/3$ & \ \ \ \ --1.25 & 1 ($u^c,e$) & \ \ \ \ --2.17\\
$e_3=2/3$ & \ \ \ \ 0.75 & 1.5 ($u^c,e$) & \ \ \ \ --2.56\\
$e_3=1$ & \ \ \ \ 4.08 & 60 ($u^c,e$) & \ \ \ \ --3.17\\
$e_8=1$ & \ \ \ \ 0.75 & 1 (non) & \ \ \ \ --0.25\\
$(m,m)$ &\ \ \ \  --0.25 & 1.5 (non) &\ \ \ \ --0.64\\
$(1,2)$&\ \ \ \ --0.59& 60 (non) & \ \ \ \ --1.25\\
\hline
\end{tabular}
\end{center}
\vskip 0.5cm

In Fig. 1, we compare the model predictions with the existing
data \cite{cav1,cav2} and the present and future sensitivities
of LLNL experiment \cite{llnl}. The experimental data are presented
with the axion number density given by Turner \cite{turner}. 
In the standard Big Bang cosmology, the axionic string and domain
walls attached to it does not give the observed cosmological
parameters if the domain wall number ($N_{DW}$) is not one \cite{dw}. 
In this case, the DFSZ model with $N_{DW}=6$ is not cosmologically
viable. Even for $N_{DW}=1$ models, the string--wall system radiate 
axions in the evolving universe. The recent estimate gives a stronger
bound on $F_a$, $F_a\le 4\times 10^{10}$ GeV \cite{shellard} than
the bound coming from cold axion density \cite{cbound}. [Note,
however, that Harari and Sikivie \cite{hara} gives a roughly the same
bound as the one from cold axion density.] 
In the inflationary cosmology, this domain wall restriction
is not applicable if the reheating temperature $T_{RH}$ after inflation
is below the axion decay constant $F_a$.
In supergravity, if the gravitino mass is around the
electroweak scale, the constraint coming from the disruption of 
nucleosynthesis from the decay products of regenerated gravitino
restricts $T_{RH}<10^{9-10}$  GeV \cite{ellis}. 
In any case, we may need an inflation with a low reheating 
temperature. Then the energy density from cold axions is the
dominant one.  The vertical axis of Fig. 1 
is $\propto c_{a\gamma\gamma}^2\times F_a^2$.
It is obvious from the figure that some models will soon confront 
serious experimental data. 
If the very light axion is not detected with the
present sensitivity of LLNL, for example the DFSZ model with
($u^c,e$) unification and the KSVZ model with $e_3=1$ and $e_Q=0$ 
are ruled out.

Before closing, we recapitulate the viability of the superstring 
axion as the solution of the strong CP problem.
If invisible axion is discovered, it cannot pinpoint which model
is correct as is obvious from Fig. 1. We regard this unpredictability
as a consequence of an ad hoc introduction of PQ symmetry.
Most probably, many heavy quarks carrying nonvanishing PQ charges
would exist, and the light quarks may also carry PQ charges.  If a
fundamental theory exists, it should predict in that framework
the invisible axion. In this regard, the discovery of
the superstring model-independent axion (MIa) is of most fundamental
importance \cite{witten}. However, the MIa decay constant is
several orders larger than the cosmological upper bound \cite{ck}.
In string models, it is known that there is no global symmetry
except the one related to a constant shift of the model
independent axion field \cite{banks}, $a_{MI}\rightarrow a_{MI}$ 
+ (constant). In other word, there is a nonlinearly realized
Peccei-Quinn symmetry in string models. We have to lower the
axion decay constant to $\sim 10^{12}$ GeV, not to violate
the cosmological energy density bound. This lowering
can be achieved \cite{kim} in 4 dimensional string models with 
an anomalous $U(1)$ gauge symmetry \cite{dsw}. This is because the 
anomalous $U(1)$ gauge boson eats up the MIa as its 
longitudinal degree of freedom \cite{kim} 
through the Green-Schwarz term \cite{green}, 
and leaves a global symmetry below this 
gauge boson mass scale. Then, this global symmetry can be
broken at the intermediate scale $\sim 10^{12}$ GeV
for example by a VEV(s) of the PQ charge carrying singlet 
scalar field. This leads to the very light (invisible) 
axion we discussed above. In general, this kind 
of model gives the contribution to $c_{a\gamma\gamma}$ 
both from the heavy quark sector and from the standard 
model quarks. If a {\it standard} superstring model is 
known, then one can calculate a unique value for $c_{a\gamma\gamma}$. 
At this moment, we do not have a standard
superstring model but can only point out
that superstring models with anomalous $U(1)$ have the room for 
the invisible axion which is on the verge of confronting data.

In conclusion, in view of the progress of axion detection
experiment, one can soon distinguish several toy models
for the invisible axion. If detected, it would open a new road
toward a fundamental theory, presumably in superstring
models.

\noindent {\it Note added}: After submission, we found that the
LLNL group actually excludes the left-hand side tip of
the sensitivity region [LLNL now] up to $m_a=3.31\times
10^{-6}$ eV [C. Hagmann {\it et al.}, LLNL preprint
astro-ph/9801286].
 
\acknowledgments
I would like to thank S. Chang for drawing the figure.
This work is supported in part by the Distinguished Scholar Exchange
Program of Korea Research Foundation and NSF-PHY 92-18167. One of
us (JEK) is also supported in part by the Hoam Foundation.

\newpage 
\begin{figure}
\epsfxsize=160mm
\centerline{\epsfbox{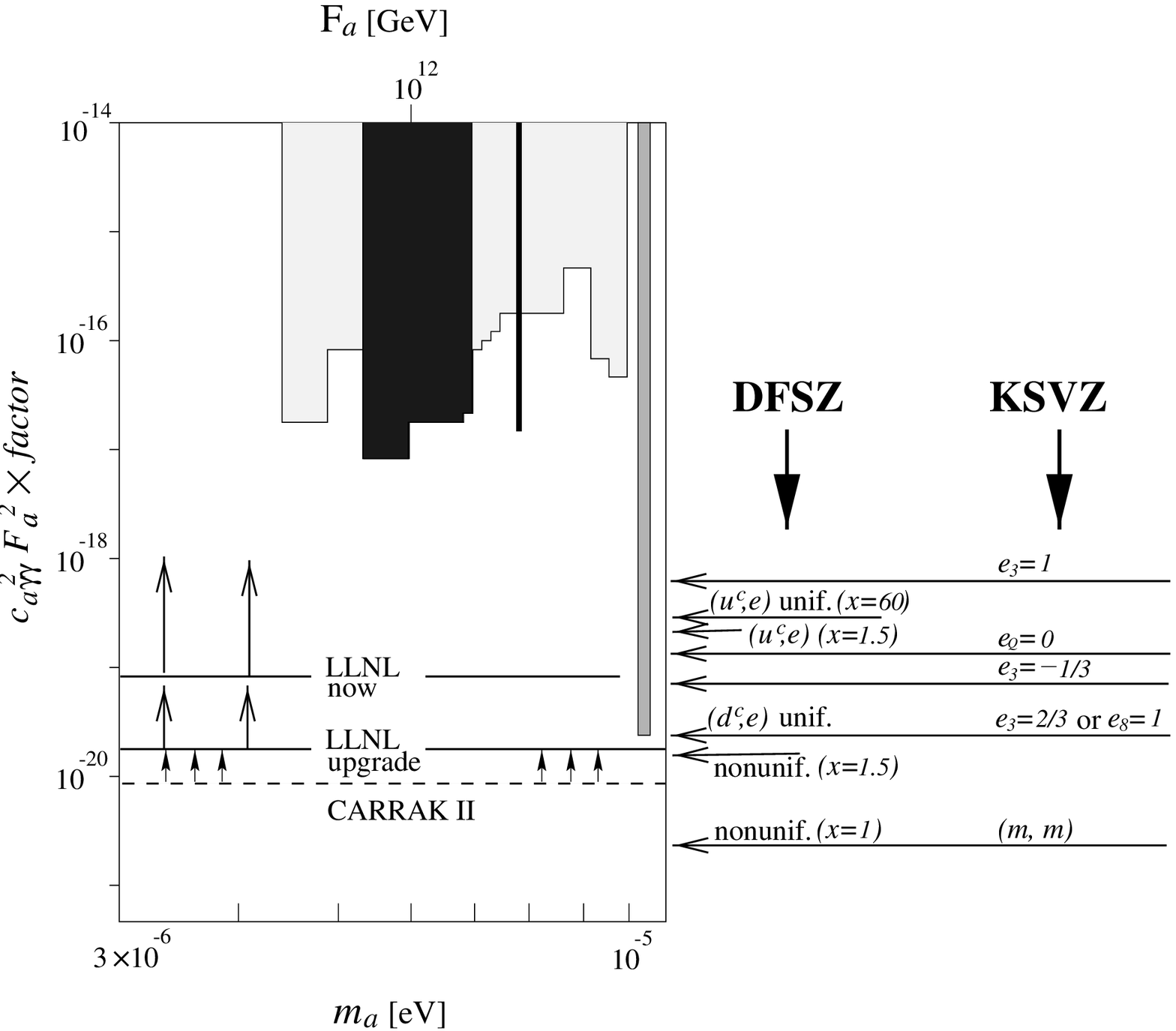}}
\end{figure}
Fig. 1. Comparison of several $c_{a\gamma\gamma}$'s 
with high-q cavity experiments. The grey region is excluded from
Rochester-Brookhaven-Fermilab experiment and the black region is
excluded from the Univ. of Florida experiment. The
present and future sensitivities of LLNL experiments are also shown.
The long column around $m_a\sim 1.0067\times 10^{-5}$ eV is the
excluded one from CARRAK I experiment \cite{matsuki}. The sensitivity
of CARRAK II is also shown.

\end{document}